# DIRECT MEASHUREMENT OF LASER ABERRATION AND AHEAD POINT FROM ARTEMIS SATELLITE THROUGH STRONG CLOUDS

Volodymyr Kuzkov
Department for space geodynamics and astrometry
Main astronomical observatory (of National Academy of Sciences of Ukraine), Kyiv, Ukraine
kuzkov@mao.kiev.ua

Sergii Kuzkov
Department for space geodynamics
Main astronomical observatory (of National Academy of Sciences of Ukraine), Kyiv, Ukraine
skuzkov@mao.kiev.ua

Serhii Borysenko
The Laboratory for Physics of Minor Solar System Bodies
Main astronomical observatory (of National Academy of Sciences of Ukraine), Kyiv, Ukraine
borisenk@mao.kiev.ua

*Abstract*— Laser communication has advances in compared with radio frequency communication as result of much high carrier frequency from ultraviolet to near infrared. Very narrow laser beam is possible to form with very high power density. But laser beam has high destruction and attenuation on clouds, turbulence, scattering on aerosols and molecules of the atmosphere. Low Earth orbits (LEO), Middling Earth orbits (MEO) and partly Geosynchronous Earth orbit (GSO) satellites moving on the sky and laser light from satellites moves across different turbulence conditions of the atmosphere, clouds, molecules of the atmosphere $H_2O$, $O_2$, $N_2$, CO, $O_3$ and other. We performed unique experiments with propagation of laser beams from beacon of OPALE terminal of ARTEMIS satellite through thin clouds. We have found that small part of laser radiation is received from ahead point there the satellite will be after time of propagation of laser radiation from the satellite to telescope. It is in accordance with theory of relativity for aberration of light during transition from moving to not moving coordinate systems. It is positive effect for laser communication through the atmosphere and clouds because will be possible to develop a system for reduce of the atmosphere turbulence during of laser communication from ground to the satellites. The interest is what will be during propagation of laser radiation from the satellite through strong clouds. The detail descriptions of laser experiment with ARTEMIS GSO satellite through strong clouds and estimations of the laser power through strong clouds are presented in this paper. Accordingly we must search the optimal wave lengths and power of lasers for performs laser communication in different cloudy conditions.

## I. INTRODUCTION

The free space laser communication systems have some advantages in comparison with the radio frequency communication systems, which is a result of much higher carrier frequency from ultraviolet to near infrared. The laser communications have a possibility of higher communication rates more than N × 10 Gbps per each communication channel.

In July 2001, the European Space Agency (ESA) geostationary Earth orbiting (GEO) Advance data-Relay and Technology Mission Satellite (ARTEMIS) was launched with on-board laser communication terminal OPALE. In 2001 the world's first laser inter-satellite communication links between ARTEMIS with on-board OPALE terminal and SPOT4 with PASTEL terminal was performed [1]. The 1789 laser communications sessions were performed between ARTEMIS and SPOT-4 (PASTEL) from 01 April 2003 to 09 January 2008 with total duration of 378 hours. Laser communication experiments between ESA's Optical Ground Station (OGS) and ARTEMIS were also performed [2,3,4]. Ground-to-satellite optical link tests between the Japanese communication terminal and the European geostationary satellite ARTEMIS has been performed [5].

The German Space Agency (DLR) and the Tesat-Spacecom designed space laser communication terminals using BPSK (binary phase shift keying) modulation and established laser communication links between LEO TerraSAR-X and NFIRE satellites (achieving data transfer rates of 5.6 Gbps at distance 5,100 km), in 2008 [6,7].

ESA is now developing the European Data Relay Satellite (EDRS) system with the use of laser communication technology to transmit data from LEO satellites to two geostationary satellites (EDRS-A and EDRS-C) with data rates of 1.8 Gbps at LEO-to-GEO link distance up to 45,000 km by using Tesat-Spacecom laser communication terminals [8].

In October 2015, NASA demonstrated transmission of data from lunar orbit LADEE space-craft to NASA OGS with a rate of 622 Mbps by using two simultaneous channels and pulsed position modulation at a distance up to 239,000 miles. The tests have been also performed for providing continuous measurements of the distance by the same laser beams from the Earth to the LADEE spacecraft with an accuracy of less than 10 mm [9].

The amount of information from telecommunication satellites in GEO constantly increases and there is a demand for a high-rate information transmission from the ground, in particular, by laser link via atmosphere. To mitigate the influence of atmospheric conditions on ground-to-space and space-to-ground laser communication, it would be desirable to have a network of OGS with different atmosphere conditions.

ESA's OGS uses the Coude focus of a 1m telescope located at an altitude of 2400 m above sea level. In 2005, the Main Astronomical Observatory (MAO) of the National Academy of Science of Ukraine started the development of a ground laser communication system for the 0.7 m AZT-2 telescope using the Cassegrain focus of a 0.7 m telescope at an altitude of 190 m above sea level. Some design works have been performed [10–12]. Therefore, it was interesting to compare the influence of atmosphere conditions in different atmosphere regions. Comparative study of atmosphere turbulence at ESA's OGS and MAO telescope has been performed as well [13–14]. Some experiments with ARTEMIS satellite started. The precision pointing and tracking system for AZT-2 telescope and other systems were developed [13, 14, 15]. As a result, the MAO developed a compact laser communication system called LACES (Laser Atmosphere Ccommunication Experiments with Satellites) [16].

## II. LASER EXPERIMENTS WITH ARTEMIS SATELLITE THROUGH THIN CLOUDS

### A. The splitting of laser beam from the satellite along declination direction

Some laser experiments with ARTEMIS satellite were performed in variable cloudy conditions at an altitude above the horizon from 22 to 25 degrees. The laser beacon beams from the satellite were recorded through thin clouds by tracking CCD camera with 732×582-pixel censor. The calculated pixel scale in the focal plane was 0.327 arc-sec per pixel for X($\alpha$) and 0.316 arc-sec per pixel for Y($\delta$) directions. The beacon transmits laser radiation at the wavelength band 797—808 nm. Clouds have an additional effect on laser radiation from the satellite. The splitting of laser beam from the satellite was observed in Y (declination) directions and X (right ascension) directions. The example of splitting is on Fig.1 and Fig.2.

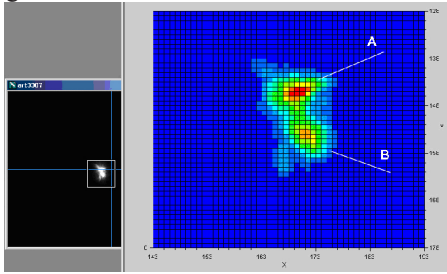

**Fig. 1. Splitting of image "art 3307".**
**The angular separation between components A and B was:**
**ΔY = 2.844″.**

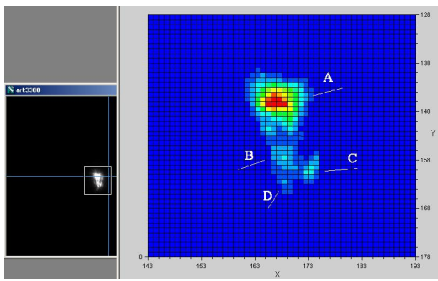

**Fig. 2. Splitting of image "art 3308".**

The angular separation was: ΔY = 3.792 arc-sec between A and B components; ΔY = 4.424 arc-sec between A and C components and ΔY= 5.372 arc-sec between A and D components.

Clouds have an additional effect on laser radiation from the satellite. Different clouds may be at different altitudes, temperatures and as result with different refraction. In addition extreme refraction is in the atmosphere at low altitudes above the horizon.

LIDAR ground systems are intended for the investigation of Raman scattering of the return reflected laser radiation on molecules in the atmosphere. They measure the nitrogen vibration-rotation Raman signals at 387 and 607 nm. The signals at 407 nm from water vapor molecules were also observed.

Raman components have frequency shift in compared with coherent laser radiation. It is negative effect. If we use very narrow interference filters and very coherent laser radiation the Raman scattering reduce the signals from and to the satellites.

Surely, Raman scattering is a very weak effect. However, in our case, due to the resonant scattering by micron-sized droplets, this mechanism could become quite realistic [17]. It is possible small direct Raman scattering of laser radiation during propagation from a satellite to ground through the atmosphere.

Rayleigh scattering on particles is less (0.1-0.2) wavelengths are smaller in intensity.

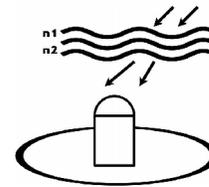

**Fig. 3. Laser beams via clouds.**

More details about splitting of laser beam along declination (meridian) direction during laser experiments from ARTEMIS through thin clouds was published in [18,19].

### B. The splitting of laser beam from the satellite along right ascension direction

The angular splitting between A and C components for image art3175 and between A and B components of image art4331 (Fig.4) was: ΔX($\alpha$) = 6 pixels or 1.962 arc-sec in the right ascension direction.

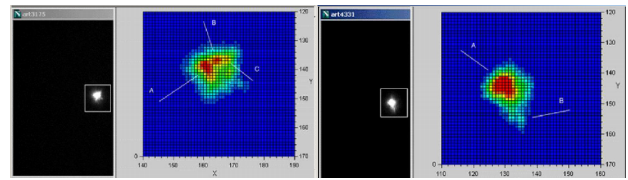

**Fig. 4. Splitting of images "art 3175" and "art4331.**

The angular splitting between A and C componentrs of the image art3308 was also ΔX($\alpha$) = 1.962 arc-sec (Fig.2b). The maximum signal for **A** component was **47579** levels of the analog to digital converter (ADC) and **19486** levels of **C** component. Signal noise was approximately **60** levels. So, for image art3308 the signal/noise ratio was **793** for **A** component and **325** for **C** component.

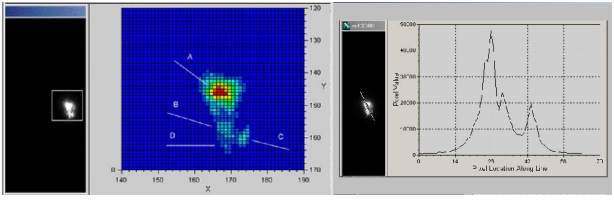

**Fig.2 b. The maximum of signals for A and C components.**

The next beacon peaks observed through clouds can be seen in image art3350 (Fig.5,a) and image art3351. (Fig.5, b).
For these images the angular splitting between A and B components was also ΔX(α) = 1.962 arc-sec.

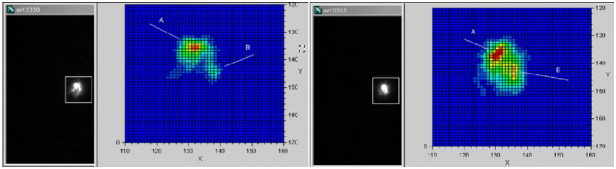

**Fig. 5. Splitting of images "art3350"(a) and "art3351" (b).**

For precision tracking of the satellite a right ascension angles (α), hour angles (H), declinations (δ) of the telescope, distances L to the satellite and time of propagation Tsig of laser radiation from the satellite to the telescope primary were calculated. Tsig = L×Vsig$^{-1}$ were Vsig = C which is the velocity of light (299,792 km×s$^{-1}$) in space.

From NORAD 2-line elements of the satellite data we know that velocity V of the satellite is 3.07 km×s$^{-1}$. The point–ahead angle Qf as shown at Fig 6a is determined as Qf = Lab × L$^{-1}$ or Qf = V × C$^{-1}$. The result is Qf = 2.112 arc sec.

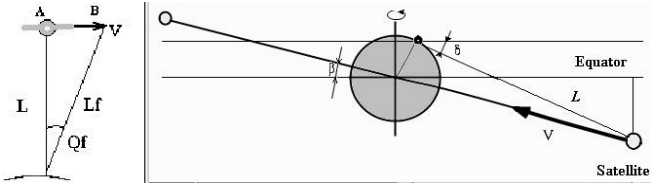

**Fig. 6 a,b. Point ahead angle left (a) and orbit of the satellite (b).**

The direction of moving satellite was from A to B (Fig.6a). The point-ahead angle Qf was calculated for orbit plane of the satellite that has inclination β = 10°20'28" to the Earth Equator plane (Fig.6 b).

Accuracy of observations was limited by size of 1 or 2 pixels and was 0.164– 0.327 arc-sec along of X axis of the CCD camera.
The accuracy of tracking the satellite was measured for image art3176 and art3229 with record time 19h:03m:03s
and 19h:06m:33s,respectively. Differences of position between observed beacon in these images were ΔX(α) = 4.532 arc-sec and ΔY(δ) =3.735 arc-sec. The drift of tracking was 0.021 arc-sec per second of time in the X(α) axis and 0.018 arc-sec per second of time in the Y(δ) axis.
As a result, a small part of the laser beam has been found to be observed ahead of the velocity vector simultaneously in the point where the satellite would arrive at for the time of propagation of laser light from the satellite to the telescope [20].

However, from (Fig. 6.a) one can see that Lf > L. Tsig for L > Tsig for Lf + T moving of satellite along L(A–B) if velocity Vsig for Lf = C velocity.
How can these observed results are explained from the stand point of modern physics?

### C. Aberration of laser light from ARTEMIS

Star aberration, as a result of moving of Earth around Sun, was discovered by James Bradley with results presented in BradleyJ. (1727) [22]. We studded the motion of ARTEMIS satellite in space around of Earth. In accordance with the theory of relativity, the aberration of light is changes the direction of light during the transition from immovable to movable coordinate systems. We have two coordinate systems of the satellite motion around of the Earth. The first one (X',Y',Z') is for the satellite. The second coordinate system (X,Y,Z) is for the ground telescope. The light direction θ' in the satellite coordinate system is determined by Eq.(1) in accordance with description presented by C.Moller (1972) [23] and assume that C much more of V we achieve:

$$\tan(\theta') = C / V \quad \ldots\ldots\ldots\ldots\ldots \quad (1)$$

When the satellite is tracked, the X axis is parallel to X' and the Y axis is parallel to Y'. The satellite center is equivalent to the center of X',Y',Z' coordinates calculated for every time of observations. The center of our telescope CCD camera is equivalent to the center of X,Y,Z coordinates. As a result, the angle θ is 90°. It is in good agreement with the calculated angle θ' determined by aberration equation (1) (V<< C). The light aberration angle (the point-ahead angle) is determined as Δθ = θ – θ'. In our case Δθ = 2.112 arc-sec (V = 3.07 km/s, C = 299,792 km/s). Thus, Δθ = Qf as calculated previously. We directly observed the laser light aberration as result of satellite moving with velocity close to standard velocity of geostationary satellites. For the time being, no other examples of direct observations of this aberration for satellites have been known. More details are presented in [20, 21].

### III. LASER EXPERIMENTS WITH ARTEMIS SATELLITE THROUGH STRONG CLOUDS

During of laser experiments with ARTEMIS on 12-15 November 2012 periods the weather conditions were unstable in Ukraine and Kyiv. The decision was to use the Ukrainian network of optical ground stations (UMOS) in different regions of Ukraine to perform synchronous test of laser active ARTEMIS satellite. Eight telescopes with objectives from 0.23m to 1.0 m were used. During 12-15 November some ground telescopes observed and registered the laser active ARTEMIS satellite [24]. Unfortunately during all days the nights was with strong clouds in Kyiv. 15 November we calculated orbit of ARTEMIS, pointed our AZT-2 telescope in calculated position and start tracking. We scanning by the telescope around possible position of ARTEMIS and recorded the images of our pointed CMOS color camera. In one image Capture-meteor-0063.jpg we detected small signals from beacon of ARTEMIS. They was presented in first time at Fig. 10, and Fig.11 of [24]. They was in some times more than noise of clouds. The detail investigations of these signals were performing below. The small part of image Capture-meteor-0063.jpg with the beacon signals is at Fig.7.

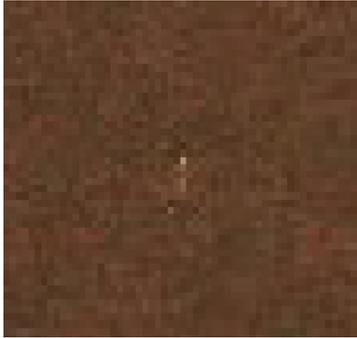

**Fig.7. Beacon laser signals of ARTEMIS through strong clouds in JPG file that is small part of image of color CMOS camera**

The first component (A) is at up. The second component (B) Is below. The third component (C) is below and left. For pointing we are using Canon color camera with CMOS sensor. The mean scale for Canon focus of AZT-2 is: $X(\delta)$ = 0.308 arc-sec per pixel,$Y(\alpha)$ = 0.303 arc-sec per pixel.
The Image Capture-meteor-0063 was recoded in jpg format on 15November 2012 during 23h:04m:08c–23h:04m:18c local time by CMOS camera Canon across of strong clouds with exposition 10 s. Binning 2×2 pixels was used for incising of sensitivity.Direction of moving the satellite is along $Y(\alpha)$ axis.Calculated orbit coordinates of ARTEMIS satellite from NORAD 2-line orbit data and from Geocentric orbit data for UTC time is below::

**From NORAD orbit data:**
```
Date      Time UTC   RA    Declin. Range Lat Long Alt   Dopp Dopp
MM/DD/YY  HH:MM:SS   Deg   Deg     Km    Deg Deg  Km    Up   Down
11/15/12  21:03:00   02:08 -11:01  38790 -3  22   35797 575  595
11/15/12  21:04:00   02:09 -10:59  38786 -3  22   35797 575  595
11/15/12  21:05:00   02:10 -10:56  38782 -3  22   35797 575  595
```
**From Geocentric orbit data:**
```
Date    UTC       RA.        Declin.   Hour Angle Alt. Moon  Speeds
2012 y  h m s     h m s      o , „     h m s      o    o     RA"/sec D
15 Nov  21:03:00  02 07 41.1 -11 01 35 00 38 42.1 28   122   0.210 2.712
15 Nov  21:04:00  02 08 40.4 -10 58 52 00 38 43.0 28   122   0.211 2.717
15 Nov  21:05:00  02 09 39.7 -10 56 09 00 38 43.8 28   123   0.213 2.721
```
Performing observation of star maps we see that no stars brighter of V 8.95 mag. in field of view 53.3x40 arc-min with satellite position at RA = 02h:08m:40.4s, Declin. = -10deg:58arc-min:52 arc-sec in the center. Previous observations in clear sky conditions indicate that laser beam from beacon of ARTEMIS observed as equivalent of 0-2 Star magnitude.

The Image Capture-meteor-0063.jpg was separated on 3 images: Capture-meteor-0063-Red, Green and Blue files. Slice of A,B,C components was performed in Red, Green and Blue colors. The results are below:

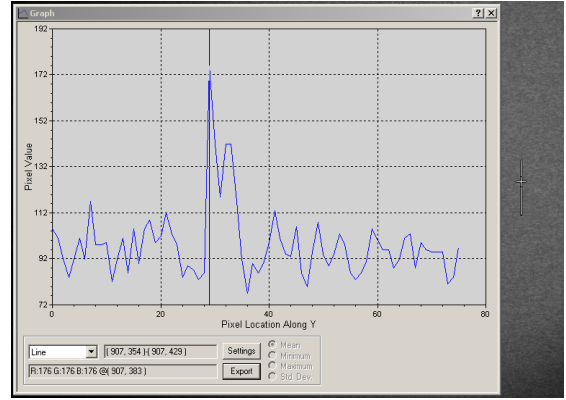

Fig.8. Capture-meteor_00623-**Red**- slice-**A**-B components. Position A component is $X(\delta)$ = **907 pixels** and $Y(\alpha)$ = **383 pixels**. Amplitude of A signal is **176 levels** of ADC (analog to digital converter) Ratio signal /noise is **176 / 117 = 1.50**.

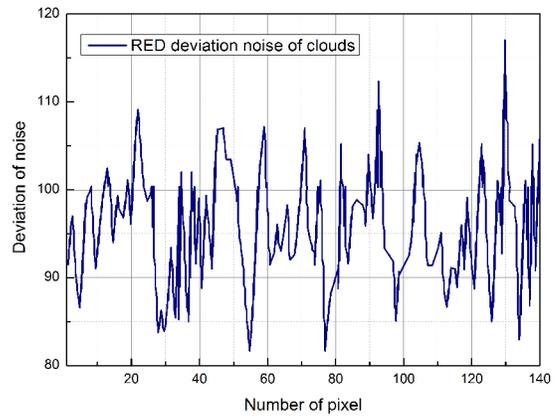

Fig.9. **Red** deviation of noise before of **A** component. Maximum short time jump of noise is **117 levels** of ADC

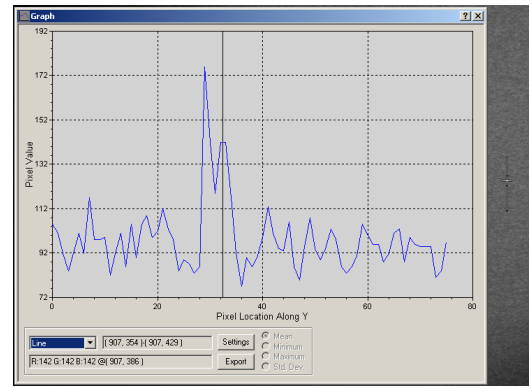

Fig.10. Capture-meteor_00623-**Red**- slice-**B**-A components. Position **B** component is $X(\delta)$ = **907 pixels** and $Y(\alpha)$ = **386 pixels**.
**ΔY(A-B) = Δ(α) = 3 pixels = 3×0.606 arc-sec per pixel = 1.818 arc sec.**
Amplitude is **142** levels of ADC. Amplitude **A/B = 176 / 142 = 1.25.**
Maximum jump of **Red** noise value is **117** levels of ADC converter.
Amplitude **Red B** component is **142** levels of ADC. Ratio **signal /noise** is **142 / 117 = 1.21.**

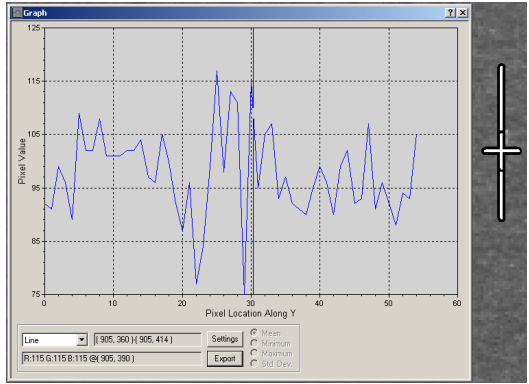

Fig.11. Capture-meteor_00623-**Red**- slice-**C** component. $X(\delta) = 905$ pixels and $Y(\alpha) = 390$ pixels. **A-C** = $\Delta X(\delta)$ =907-905 = 2 pixels = **1.232 arc-sec. A-C** = $\Delta Y(\alpha)$ =390 – 383 = 7 pixels = **4, 312 arc-sec .** Amplitude. is comparative with noise of clouds and is approximately **95** levels of ADC.

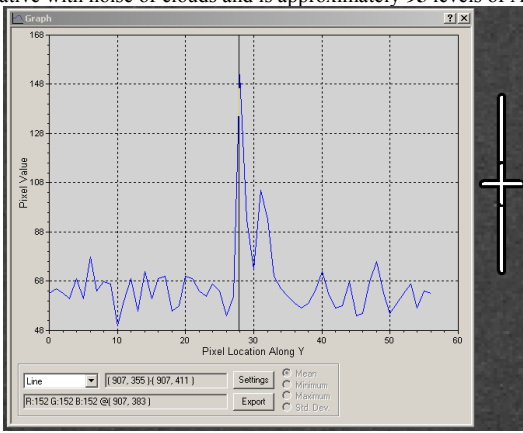

Fig.12. Capture-meteor_00623-**Green**-slice-**A**,B components. Position of A component is $X(\delta) = 907$ pixels and $Y(\alpha) = 383$ pixels and has same positions as Red –slice-A-B components. Amplitude is **152** levels of ADC.

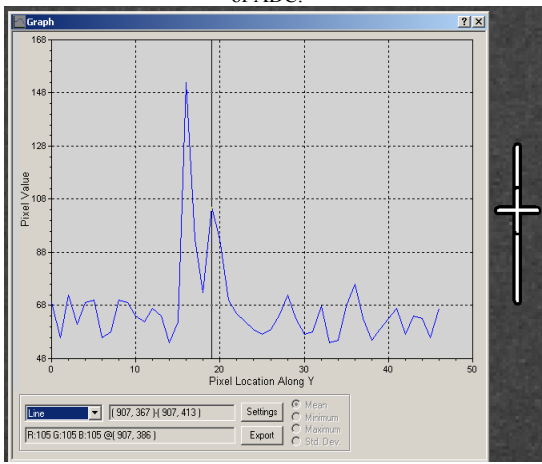

Fig.13. Capture-meteor_00623-**Green**-slice-**B**-A components. Position B component is $X(\delta) = 907$ pixels and $Y(\alpha) = 386$ pixels and same positions as **Red** –slice **B**-A components. Amplitude is **105** levels of ADC. Amplitude **A / B** = **152/105** = **1.45.** $\Delta Y(A\text{-}B)$ = $\Delta(\alpha) = 3$ pixels = $3 \times 0.606$ arc-sec per pixel = **1. 818 arc-sec.**

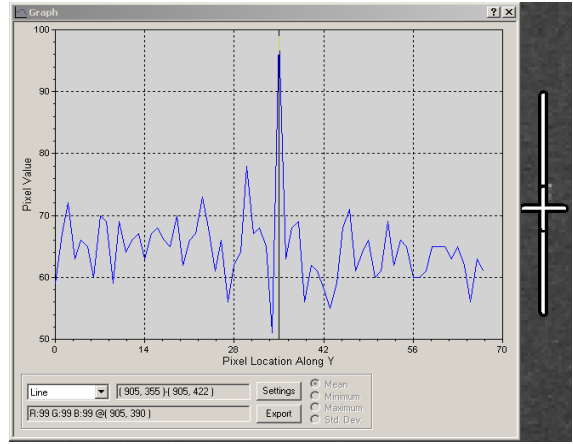

Fig.14. Capture-meteor_00623-**Green**-slice-**C**-components. Position C component is $X(\delta) = 905$ pixels and $Y(\alpha) = 390$ pixels. Amplitude is **99** levels of ADC.

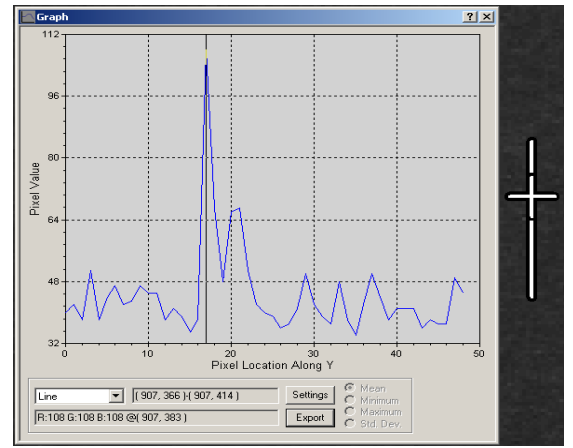

Fig.15. Capture-meteor_00623-**Blue**-slice **A**-B components. Position of **A** component is $X(\delta) = 907$ pixels and $Y(\alpha) = 383$ pixels and same position as Red –slice **B-A** components. Amplitude is **108** levels AGC.

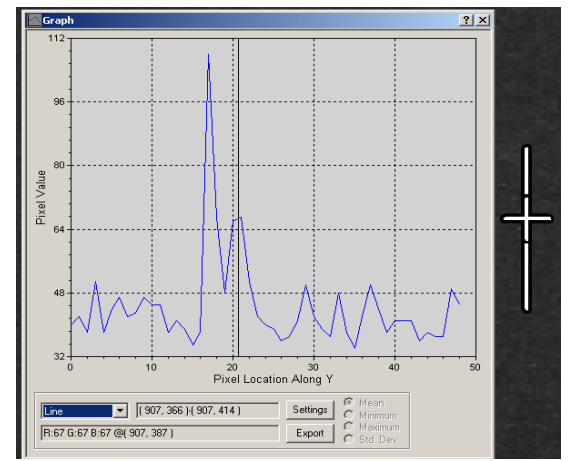

Fig.16. Capture-meteor_00623-**Blue**-slice-**B**-A- components. Position **B** component is $X(\delta) = 907$ pixels and $Y(\alpha) = 387$ pixels. $\Delta Y(\alpha)$ **A –B** = 4 pixels = **2.424** arc-sec . Amplitude is **67** levels of ADC. Amplitude **A/B** = **108 / 67** = **1. 61**.

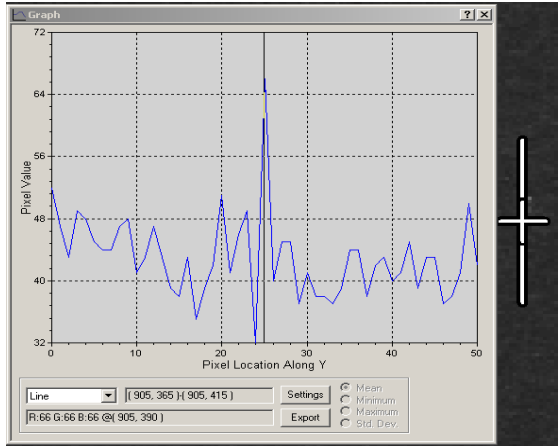

Fig.17. Capture-meteor_00623-**Blue**-slice-**C**- component. Position **C** component is X(δ) = **905** pixels and Y(α) = **390** pixels and same as **Green C** component. Amplitude is **66** levels of ADC.

Maximum jump of short time **Red** noise value is **117** levels of ADC. Amplitude **Red A** component is **176** levels of ADC. Signal **A** / Noise ratio **176 / 117** = **1. 5.**
This is a significant excess of the laser signal over the short time maximum fluctuation of the noise of the sky.

Difference for Red and Green A-B components positions is ΔY(A-B) = Δ(α) = 3-4 pixels = (3-4)×0.606 arc-sec /pixel = **1.818 – 2.424 arc –sec**. Mean velue of Δ(α) = **2.121 arc-sec**. It is close to **Qf = 2.112 arc-sec** obseved for ARTEMIS through thin clouds. Δ **Qf = 2.121– 2.112 =0.009 arc-sec** !
From NORAD orbit data we know that **V sat.** = was **3.074 – 3.076 km/s.** Point ahead angle is determited as Qf = V × C$^{-1}$.
Then calculated ahead angles are: **Qf = 2.115–2.116 arc-sec**. **2.121 – 2.115 = 0.006 arc-sec. 2.121 – 2.116 = 0.005 arc-sec.**
We also obseved aberration of laser beam from the beacon of ARTEMIS and ahead angle through strong clouds !

Summary of results of Red, Green, Blue slices for A,B,C components are below:

**Red** components:
Red - **A** component: Amplitude is **176** levels of ADC.
Red - **B** component: Amplitude is **142** levels of ADC
Red - **C** component: Amplitude is **95** levels of ADC
**Green** components:
Green-**A** component: Amplitude is **152** levels of ADC
Green-**B** component: Amplitude is **105** levels of ADC
Green-**C** component: Amplitude is **99** levels of ADC
**Blue** components:
Blue- **A** component: Amplitude is **108** levels of ADC
Blue- **B** component: Amplitude is **67** levels of ADC
Blue- **C** component: Amplitude is **66** levels of ADC

Graphics of these data presented at Fig.18.

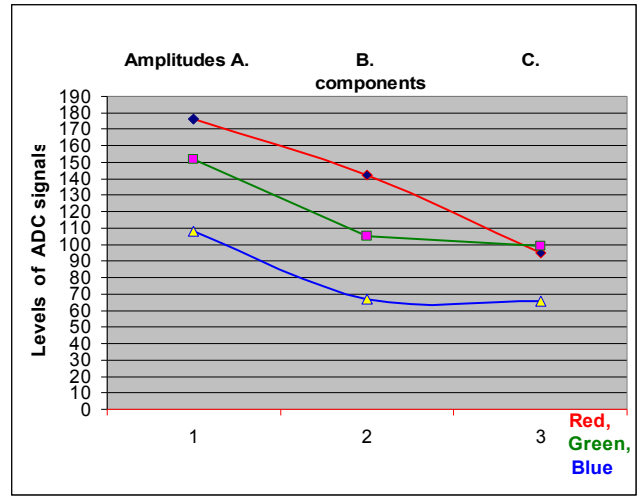

Fig.18. Graphics of **Red, Green, Blue** amplitudes for **A,B,C** components.

A, B components are much more Red and Green than Blue.
C component is more Green then Red and Blue.

## CONCLUSIONS

We performed uniquely experiments of transition laser radiation through thin and strong clouds. We found that splitting of laser beam is in meridian direction as result of different temperature, density and refraction of clouds.

Raman scattering is a very weak effect. However, in our case, due to the resonant scattering by micron-sized droplets, this mechanism could become quite realistic. Before we found that part of laser radiation is from ahead points there the satellite will be during propagation of laser radiation from satellite to ground station and direct laser beam aberration observed as result of moving of the satellite with velocity ~3.07km/s. The same ahead point we observed during of propagation laser radiation through strong clouds. Distance between Red and Green A-B components positions was ΔY(A-B) = Δ(α) = 3–4 pixels = **1.818 – 2.424 arc-sec** and is close to our measurement of ahead points for beacon of ARTEMIS that was observed through thin clouds !
Laser signal /maximum short time jump of noise of strong clouds for **Red** component was **1. 5**.
For laser communications through of Earth atmosphere main tasks are:
1. Reduce possible splitting of laser beam by different clouds with different altitudes, refraction and Raman scattering.
2. Increasing variable power of laser beam for transmitting through strong clouds.
3. The ahead angle and ahead points are forming at a satellite for optimal working of ground turbulence compensation systems.
4. Select optimal wavelength of lasers for minimize of scattering of laser radiation on molecules of the atmosphere.

## ACKNOWLEDGMENT

The authors would like to thank Zoran Sodnik from ESTEC ESA for great support in development our ground laser communication system, Redu Space Services S.A., ESA, Redu, Belgium for great assistance in preparing and performing the observations. We also would like to thank Prof. Ya. S. Yatskiv of the National Academy of Sciences and National Space Agency of Ukraine for the support in the development of our optical ground system.